\documentclass[conference]{IEEEtran}
\usepackage{cite}
\usepackage{amsmath,amssymb,amsfonts}
\usepackage{algorithmic}
\usepackage{textcomp}
\usepackage{hyperref}        % doi links in bib
\usepackage{graphicx}        % images
\usepackage{subcaption}      % sub images
\IEEEoverridecommandlockouts
\usepackage{tikz}            
\newcommand\copyrighttext{
  \footnotesize \textcopyright 2018 IEEE. Personal use of this material is permitted. Permission from IEEE must be obtained for all other uses. DOI: \href{https://doi.org/10.1109/EuroSPW.2018.00013}{10.1109/EuroSPW.2018.00013}}
\newcommand\copyrightnotice{
\begin{tikzpicture}[remember picture,overlay]
\node[anchor=south,yshift=10pt] at (current page.south) {\fbox{\parbox{\dimexpr\textwidth-\fboxsep-\fboxrule\relax}{\copyrighttext}}};
\end{tikzpicture}
}

\DeclareCaptionLabelSeparator{periodspace}{.\quad}
\begin{document}

%----------------------------------------------------------------------------------------
\title{The Impact of Uncle Rewards on Selfish Mining in Ethereum}
%----------------------------------------------------------------------------------------

\author{
\IEEEauthorblockN{Fabian Ritz\IEEEauthorrefmark{1} and Alf Zugenmaier\IEEEauthorrefmark{1}\IEEEauthorrefmark{2}}\\
\IEEEauthorblockA{\IEEEauthorrefmark{1}Munich University of Applied Sciences, Munich, Germany}
\IEEEauthorblockA{\IEEEauthorrefmark{2}Concordia University, Montreal, Canada\\
\{fabian.ritz, alf.zugenmaier\}@hm.edu}
}
\maketitle
\copyrightnotice

%----------------------------------------------------------------------------------------
\begin{abstract}
%----------------------------------------------------------------------------------------
Many of today's crypto currencies use blockchains as decentralized ledgers and secure them with proof of work. In case of a fork of the chain, Bitcoin's rule for achieving consensus is selecting the longest chain and discarding the other chain as \emph{stale}. It has been demonstrated that this consensus rule has a weakness against \emph{selfish mining} in which the selfish miner exploits the variance in block generation by partially withholding blocks. In Ethereum, however, under certain conditions stale blocks don't have to be discarded but can be referenced from the main chain as \emph{uncle} blocks yielding a partial reward. This concept limits the impact of network delays on the expected revenue for miners. But the concept also reduces the risk for a selfish miner to gain no rewards from withholding a freshly minted block.

This paper uses a Monte Carlo simulation to quantify the effect of uncle blocks both to the profitability of selfish mining and the blockchain's security in Ethereum (ETH). A brief outlook about a recent Ethereum Classic (ETC) improvement proposal that weighs uncle blocks during the selection of the main chain will be given.
\end{abstract}

\begin{IEEEkeywords}
Blockchain, GHOST, Ethereum, Uncle Blocks, Selfish Mining
\end{IEEEkeywords}

%----------------------------------------------------------------------------------------
\section{Introduction}\label{sec:intro}
%----------------------------------------------------------------------------------------
Since the release of Bitcoin in 2009~\cite{bitcoin}, many blockchain-based crypto currencies have followed. Currently, Ethereum is one of the most successful. It claims to be a ``Secure Decentralised Generalised Transaction Ledger''~\cite{ethereum-yellowpaper}. It differs from Bitcoin in that it offers a more flexible way of handling contracts, has a higher block generation rate and different mining rewards. Ethereum relies on a proof of work secured blockchain\footnote{At the time of writing, there are proposals to migrate from proof of work to proof of stake.}. Because there is money involved, miners that perform proof of work may try to devise strategies to increase their mining revenue beyond their fair share. In the literature, such a technique has been described as selfish mining. This strategy is exploiting the variance in block generation by partially withholding information~\cite{bitcoin-selfish-mining}, causing altruistic miners to waste computational resources. For a selfish miner in Bitcoin, withholding a freshly minted block from the blockchain involves a risk: by the time the selfish miner releases the block, the other miners may have advanced in the chain and this block becomes stale, thus not yielding any reward for the selfish miner. Normally, stale blocks come into existence due to propagation delays. Block generation rate and information propagation in the underlying peer-to-peer-networks have already been analyzed in depth, e.g. by Decker and Wattenhofer~\cite{information-propagation-bitcoin}. Based on that, Gervais et al.~\cite{security-performance-blockchain} quantified trade-offs between network throughput and blockchain security. Since optimized values for block size and block generation rate alone could not solve the transaction throughput limits of the existing consensus rules, more advanced variants were developed. A well-known example is the GHOST-rule~\cite{secure-high-tps-bitcoin}~\cite{trees-chains-tps-blockchain} which utilizes stale blocks in the calculation of the heaviest chain which is considered to be the main chain. To counter the negative impact of propagation delays on mining rewards in Ethereum, stale blocks may be included in the main chain as uncle blocks. Altering GHOST, these uncle blocks yield a partial reward for all involved parties. Until Ethereum's Byzantinum hard fork in October 2017~\cite{EIP-100}, this could be exploited by a technique called uncle mining~\cite{uncle-mining}. Even after this hard fork, uncle block rewards still influence the profitability of the original selfish mining strategy\cite{bitcoin-selfish-mining}. So far, there hasn't been a study to quantify this effect as well as the blockchain's remaining resilience to further attacks.

This paper shows that partially rewarding uncle blocks as done in Ethereum lowers the threshold at which selfish mining becomes profitable from $0.25$ if there are no uncle block rewards to approximately $0.185 \pm 0.012$ at an honest network's uncle block ratio as observed in December 2017. Furthermore, it also quantifies how the presence of a selfish miner weakens Ethereum against further attackers.

%----------------------------------------------------------------------------------------
\section{Background}\label{sec:background}
%----------------------------------------------------------------------------------------
As stated previously, Ethereum builds upon a public ledger to record all state changes in a sequence of blocks called \emph{blockchain}. Pending state changes are broadcast via messages between all nodes of a Kademlia based peer-to-peer-network and are eventually processed into a block's \emph{payload} by specialized \emph{miner} nodes. If a miner node succeeds to solve a cryptographic-puzzle by finding a low hash value (performing proof-of-work), the solution is propagated throughout the network. The miner receives a block reward paid in Ether (ETH) for doing so. The blockchain's \emph{height} increases by one as the new block is replicated across all nodes. The details of the Ethereum system are documented in a regularly updated whitepaper~\cite{ethereum-whitepaper} as well as a yellowpaper~\cite{ethereum-yellowpaper}.

In contrast to other crypto currencies, Ethereum features a comparably small average time between block generations -- approximately 15 seconds~\cite{etherscan-blocktime} at the time of the writing. If more than one block of the same blockchain height is propagated throughout the network simultaneously, these blocks form a \emph{fork}. Employing what is called \emph{random tie breaking}, miners randomly choose upon which fork they try to build their next block. Over time, one of the forked chains will grow longer than the other. Following the consensus protocol, all nodes will recognize the longest chain as the main chain~\cite{longest-chain-rule} and continue to extend this one. Therefore the smaller average block time of Ethereum results in a significantly higher rate of blocks not increasing the blockchain height: in 2017 it varied between 0.06 and 0.24 and reached its peak in Januar 2018~\cite{etherscan-uncles}. This peak was caused by a global hype of crypto currencies resulting in a high number of transactions~\cite{etherscan-transactions} within the Ethereum network. Therefore bigger blocks~\cite{etherscan-blocksize} were created which take longer both to propagate within the network and to be validated by the nodes.

\begin{figure}[h]
\includegraphics[width=.485\textwidth]{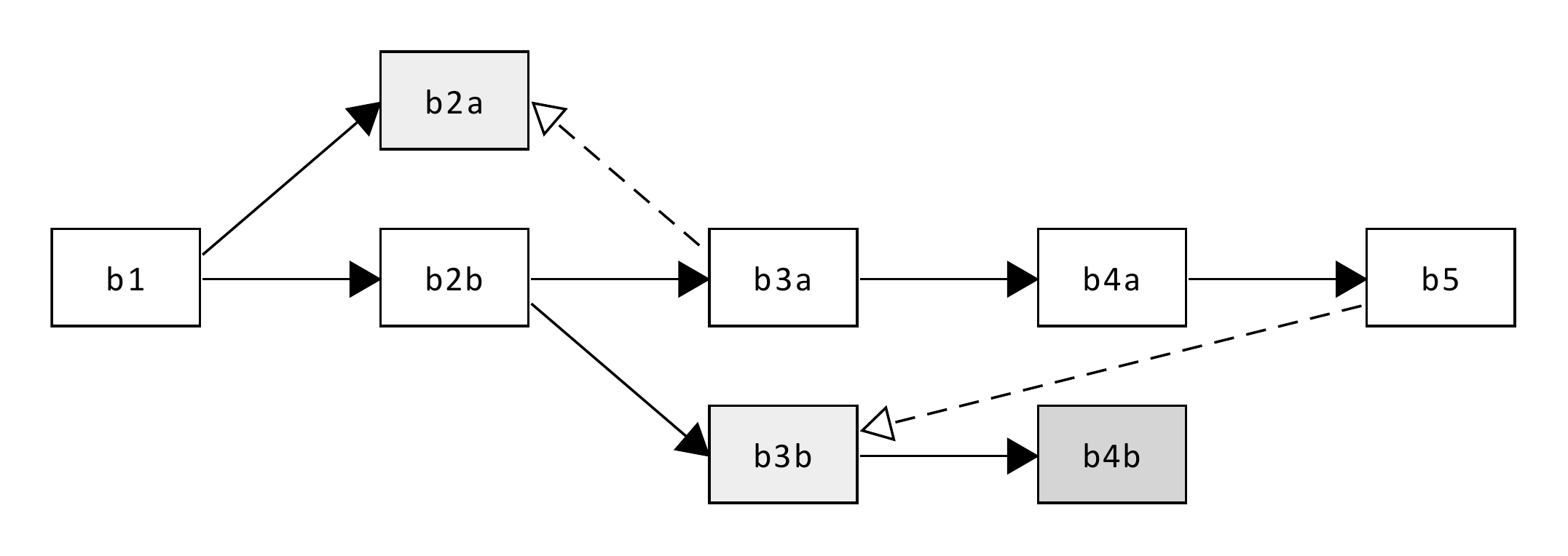}
\caption{Ethereum blockchain sample. Block b2a is a first generation uncle block due to the inclusion by block b3a with distance one. Block b3b is a second generation uncle block. Block b4b is a stale block and may not be referenced as an uncle block since it's parent, block b3b, is not part of the main chain.}
\label{fig:block-exmaples}
\end{figure}

In general, every block that is not part of the main chain is called a \emph{stale} block. Stale blocks do not yield a reward and their payload is ignored. In Ethereum, every stale block that is a direct descendant of a main chain (\emph{regular}) block may be referenced as an \emph{uncle} or \emph{ommer} block from one of the following regular blocks (\emph{nephew}) up to a maximum distance of 6 blocks after the fork (see Fig. \ref{fig:block-exmaples} with description below). Therefore uncle blocks contain a valid header. Nevertheless their payload is ignored. Up to two uncle blocks may be referenced by one regular block. The miner of the referencing block gets an inclusion reward of $\frac{1}{32}$ block reward. The uncle block miner's reward depends on how many blocks later the uncle block was referenced: the reward ranges from a $\frac{7}{8}$ block reward for an immediate inclusion to a $\frac{2}{8}$ block reward for inclusion 6 blocks later. In November 2017, an average uncle block reward of 0.725 was observed~\cite{etherscan-uncles}~\cite{etherscan-blocks}. Thus an uncle block is included by a nephew block on average two blocks later. In case of forks longer than one block, any further descendants may not be referenced as uncle blocks.

The Ethereum network tries to maintain a constant average block time. This is achieved by a dynamic adjustment of the proof-of-work difficulty according to the actual block generation rate. Formerly, only regular blocks were used to determine the network's block generation rate. This made an exploit called \emph{uncle mining}~\cite{uncle-mining} possible. Since the \emph{Byzantinum} hard fork in October 2017~\cite{EIP-100}, uncle blocks are also counted within the calculation of the block generation rate.

%----------------------------------------------------------------------------------------
\section{Model}
%----------------------------------------------------------------------------------------
%----------------------------------------------------------------------------------------
\subsection{Quantities of interest}\label{sec:quantities}
%----------------------------------------------------------------------------------------
In \emph{selfish mining}, a mining party within the network is trying to receive a share of the revenue larger than its share of mining power. Following the model of Eyal and Sirer~\cite{bitcoin-selfish-mining}, the selfish miner will not necessarily comply with the Ethereum protocol. It is assumed that his connection to the network is perfect. Therefore, he is capable of instantly receiving information about events such as block creations, and may react in time. His fraction of the total network's mining power is defined as $\alpha$. All other, honest participants are assumed to follow the Ethereum protocol as intended. Their fraction of mining power is defined as $\beta$. In the absence of a further attacker, $\alpha + \beta = 1$. In case of ties, i.e. chains of same length, Ethereum uses \emph{random tie breaking} in which miners are expected to randomly select on which block to continue mining if they notice a fork. Therefore the honest network's mining power will be diverted with a fraction of $\gamma \cdot \beta$ mining on any of the competing chains. This deviates from the original selfish mining~\cite{bitcoin-selfish-mining} as the selfish miner is not capable of influencing $\gamma$ without further attacking the honest network, e.g. with eclipse attacks \cite{stubborn-mining}, which is outside this paper's scope. This results in:
\begin{equation}\label{eq:gamma}
\gamma=\frac{1}{\textrm{number of competing chains}}
\end{equation}
Extending selfish mining as proposed in~\cite{bitcoin-selfish-mining}, the ratio of the honest network's stale block generation to honest networks overall block generation is given by $\delta$, with $\delta \le \beta$. The selfish miner is assumed to be perfectly connected, therefore will never create stale blocks within his secret fork and build a linear chain.

As in Eyal and Sirers' model~\cite{bitcoin-selfish-mining}, the selfish miner aims to increase his revenue in relation to the honest network's revenue, which is defined as the \emph{relative revenue ratio} ($\mathit{RRR}$). Both uncle blocks and regular blocks are counted within the difficulty calculation. Thus the sum of uncle block rate and regular block rate is constant. As stated previously, in Ethereum uncle blocks yield lower rewards than regular blocks. Therefore the uncle block ratio is assumed to influence the profitability of selfish mining as it causes a high uncle block ratio which lowers the overall reward. It can be assumed a selfish miner prefers maximizing his absolute revenue, which is reflected by the \emph{absolute revenue ratio} ($\mathit{ARR}$). Let
\begin{itemize}
\item $nb(p,r)$ the number of a party's regular blocks,
\item $nb(p,u)$ the number of a party's uncle blocks,
\item $nb(a,r)$ the number of all regular blocks,
\item $nb(a,u)$ the number of all uncle blocks and
\item $rev(p)$ the party's normalized total revenue from all blocks with the reward of a regular block set to 1.
\end{itemize}
$\mathit{ARR}$ may then be defined as follows:
\begin{equation}\label{eq:ARR-complete}
\mathit{ARR}=\frac{nb(p,r) + nb(p,u)}{nb(a,r) + nb(a,u)} \cdot \frac{rev(p)}{nb(p,r) + nb(p,u)}
\end{equation}
The left fraction reflects the ratio of a party's blocks that yield any reward in relation to all blocks yielding any reward. The right fraction reflects the party's average reward per block. The equation can be further simplified:
\begin{equation}\label{eq:ARR-simple}
\mathit{ARR}=\frac{rev(p)}{nb(a,r) + nb(a,u)}
\end{equation}

While absolute and relative revenue ratio measure profitability of selfish mining as such, selfish mining might also affect the resilience of the network against attacks that rely on (partial) control of the blockchain, e.g. double spending or a $0.5+\epsilon$ (colloquial 51\%) attack. Two scenarios need to be distinguished:
\begin{itemize}
\item The selfish miner could attack the network. Even if he does not earn money from the attack directly, he might be bribed by others to do so. The effectiveness of the attack primarily depends of the ratio of the selfish miner's regular blocks that extend the blockchain's height in comparison to all other blocks within the main chain -- referred to as the \emph{regular block ratio} ($\mathit{RBR}$).
\item Not the selfish miner, but another party could attack the network. As mentioned before, stale blocks do not extend the longest chain. Thus every stale block lowers the overall resilience of the blockchain against such attacks compared to a best case linear blockchain with the given mining resources. This is reflected by the \emph{relative network security} ($\mathit{RNS}$) which is defined as the number of regular blocks in relation to all blocks (including stale blocks and uncle blocks).
\end{itemize}
$\mathit{RNS}$ also outlines the network's liveliness: Since all stale blocks' payload is ignored and the stale blocks' counterparts at same height within the main chain do not necessarily contain the same payload, the missing payload needs to be added to the blockchain again within one of the following regular blocks. While the blockchain is under load, i.e. each block's payload is fully utilized, uncle blocks may have a significant impact on the network transaction throughput since uncle blocks are counted as regular blocks within the block generation rate\cite{EIP-100}.

%----------------------------------------------------------------------------------------
\subsection{Simulator}\label{sec:simulator}
%----------------------------------------------------------------------------------------
To quantify the aforementioned values of interest, attempts were made to define an MDP extending Eyal and Sirers' previous work~\cite{bitcoin-selfish-mining}. But random tie breaking and uncle block inclusion could not be expressed in closed form without various assumptions sacrificing accuracy. Consequently, a Monte Carlo Simulation was designed instead. It runs a defined number of steps within a discrete system modeling block generation events. The underlying data structure captures all relevant aspects of a blockchain and is evaluated after all blocks have been created. In each simulation step, a new block is mined -- by either the selfish miner ($\alpha$) or the honest network ($\beta_r, \beta_s$) according to the following rules:
\begin{itemize}
\item [$\alpha$] The selfish miner mines a new block on top of the best block known to him. If he already maintains a secret fork, the best block is always the top of that, otherwise the best block is the top of the longest public chain. In case of competing chains, one is chosen randomly. One way or the other, the block is kept secret for the moment. Only in case of competing chains with one having been mined by the selfish miner himself earlier, that one is selected in particular instead of a random choice. And only in that case, the block is published immediately. This event occurs with probability $\alpha$.
\item [$\beta_r$] The honest network mines a regular block on top of the best block known to the public. In most cases, this is the longest public chain. If there are competing chains, one is chosen randomly (random tie breaking). The block is always published immediately. This event is occurs with probability $1 - \alpha - (\delta \cdot \beta)$.
\item [$\beta_s$] The honest network mines a stale block competing with the best block known to the public. If there is more than one option, one is chosen randomly. This either creates a new fork out of any chains at this height or extends a fork formerly ending at this height. The block is always published immediately. This event occurs with probability $\delta \cdot \beta$.
\end{itemize}
If the selfish miner maintains a secret fork, there are the following possibilities: 
\begin{itemize}
\item The selfish miner mines another new block without the honest network extending the public longest chain. He therefore has a lead of two (or more) blocks. He will keep extending his secret fork until the honest network manages to catch up, i.e. the honest network builds a chain with a length of one block less than the selfish miner's secret fork. Then the selfish miner broadcasts his entire secret fork to have the honest network adopt his fork as the main chain.
\item The honest network catches up before the selfish miner adds a second block to his secret fork. He then publishes his secret block immediately to cause a race between his block and the one of the honest network. Because of random tie breaking, the honest network will split up equally, partially mining upon on the selfish miner's fork. Deviating from random tie breaking, the selfish miner will only mine upon his formerly secret block. If he mines the next block, he will again publish it immediately to have his blocks win the race. If the honest network mines the next block, the selfish miner will adopt it regardless of whether it was build upon his formerly secret block or not.
\end{itemize}
The described behavior matches the original selfish mining strategy~\cite{bitcoin-selfish-mining} with the selfish miner's influence on the honest network $\gamma$ being replaced by random tie breaking and the addition of honest network's stale block generation ratio $\delta$. Due to these changes, the simulator is able to emulate a network of altruistic miners in which stale blocks are created with a certain ratio, e.g. due to network latency, even in absence of a selfish miner. If a selfish miner is present, his actions may of course further alter the honest network's stale block ratio. However, these changes lead to some consequences:
\begin{itemize}
\item Every time the honest network creates a stale block, it fails to shorten the lead of the selfish miner's secret fork and  intuitively seems to give him some advantage.
\item The honest network might not always adopt a selfish miner's fork but extend either a shorter fork that was previously outrun by one block or even fork the selfish miner's fork. This causes a race with uncertain outcome for the selfish miner and intuitively seems to give him some disadvantage.
\item The selfish miner may choose to build the next secret block on what will afterwards become a stale block. Assuming he then loses a race to break a tie against the honest network, his block can not be included as an uncle block. He might however win the race or even increase his lead to two (or more) and outrun the honest network anyway.
\end{itemize}
As mentioned in Section \ref{sec:background}, the Ethereum blockchain allows stale blocks to be referenced as uncles under certain conditions. It forms a tree-like structure with each block having been placed at its position with the intention of being or becoming part of the main chain. Based on that, each particular block's possible uncles can only be first generation descendants of any forks on the ascending path to the genesis block as seen from this particular block. To maximize the expected reward, more distant blocks are preferred -- nearer blocks might still be referenced from the current block's descendants while more distant blocks may then be out of range -- to overall include as many uncles as possible. The honest network is assumed to always include any blocks as uncles while a selfish miner might chose to include any blocks, own blocks or no blocks at all. Since on the current Ethereum blockchain, the average uncle block is included on average two blocks away (see Section \ref{sec:background}), the honest network's chance to actually include an available uncle block was set to 0.33 per uncle block slot and simulation step. In simulations without a selfish miner, this resulted in the same average uncle block reward as observed in the Ethereum network in November 2017 (see Section \ref{sec:background}). In contrast, the selfish miner's chance to include an uncle block is always 1.0 -- justified by the assumption of a perfect network connection. If a selfish miner is present, his actions may of course alter the average uncle block reward in the honest part of the network further.

%----------------------------------------------------------------------------------------
\subsection{Simulation}\label{sec:simulation}
%----------------------------------------------------------------------------------------
The previously outlined Monte Carlo Simulator was implemented to quantify the values of interest depending on the selfish miner's percentage of computational power. For each parameter combination of $\alpha$, selfish mining strategy, $\delta$ and uncle block inclusion strategy, 100 random walks with at least $2^{17}$ generated blocks were run and analyzed to achieve precise mean values. Utilizing more random walks with fewer generated blocks per walk turned out to be less precise because of border effects, e.g. the treatment of an unpublished fork at the end of an simulation, have bigger impact and occur more often. For parameters near the break even point, i.e. $0.15\leq\alpha \leq0.30$, each mean value was only accepted if the underlying results' standard deviation $\sigma$ was smaller than a threshold $T_1$ of 0.001. This more precise threshold applied in the range in which we expected the break even point. If $\sigma$ was greater than $T_1$, the simulation was re-run with twice as many generated blocks until the standard deviation was smaller than $T_1$. Using this value of $T_1$, it is possible to derive from the simulation results that the standard deviation of the $\alpha$-value of the break even point is 0.012.  For any other values of $\alpha$, a higher threshold $T_2$ of 0.01 was defined, accepting a possible loss of accuracy.

To check plausibility, the results were verified against known reference values. Since selfish mining has not been analyzed in Ethereum yet, but is a well covered topic in Bitcoin, the simulator was first adjusted to simulate Bitcoin blockchains: the honest network's stale block generation ratio $\delta$ was set to zero, the selfish miner's and the honest network's uncle block inclusion strategy were set to \emph{none}. This ensured that stale blocks would never be referenced as uncles and only occur in presence of a selfish miner temporary forking the honest network's main chain. Consequently there would never be more than two conflicting best blocks. Due to random tie breaking, this causes the honest network to behave exactly as the model of Eyal and Sirer~\cite{bitcoin-selfish-mining} with the attacker's network influence $\gamma$ set to 0.5. With this configuration, the simulator was able to reproduce the corresponding results of Eyal and Sirer~\cite{bitcoin-selfish-mining}.

While it was intended to compare the simulation results with experimental results within a private testnet, the client of the Ethereum Java reference implementation \textit{ethereumj}~\cite{ethereumj} did not reliably switch to the longest chain. These synchronization issues affected solely the mining nodes if two or more mining nodes were present and could not be solved during the work for this paper. Thus, ethereumj couldn't be used to verify mining strategies. This bug seems to be low priority, as a single miner node suffices to sustain a private network to experiment with Ethereum except for the exploration of different mining strategies. Mining in the main network is backed by OpenCL- or CUDA-clients such as \textit{ethminer}~\cite{ethminer} utilizing graphics cards due to their much higher computing power.

%----------------------------------------------------------------------------------------
\section{Results}\label{sec:results}
%----------------------------------------------------------------------------------------
The following Figures \ref{fig:RRR}, \ref{fig:ARR} and \ref{fig:RBR-RNS} depict the results of a Monte Carlo Simulation as defined in Section \ref{sec:simulator}, having been run as described in Section \ref{sec:simulation} under the assumptions of Section \ref{sec:quantities}.

\begin{figure*}
\centering
\begin{subfigure}{\columnwidth}
\includegraphics[trim=2cm 1.5cm 2cm 1.5cm, clip=true, width=\columnwidth]{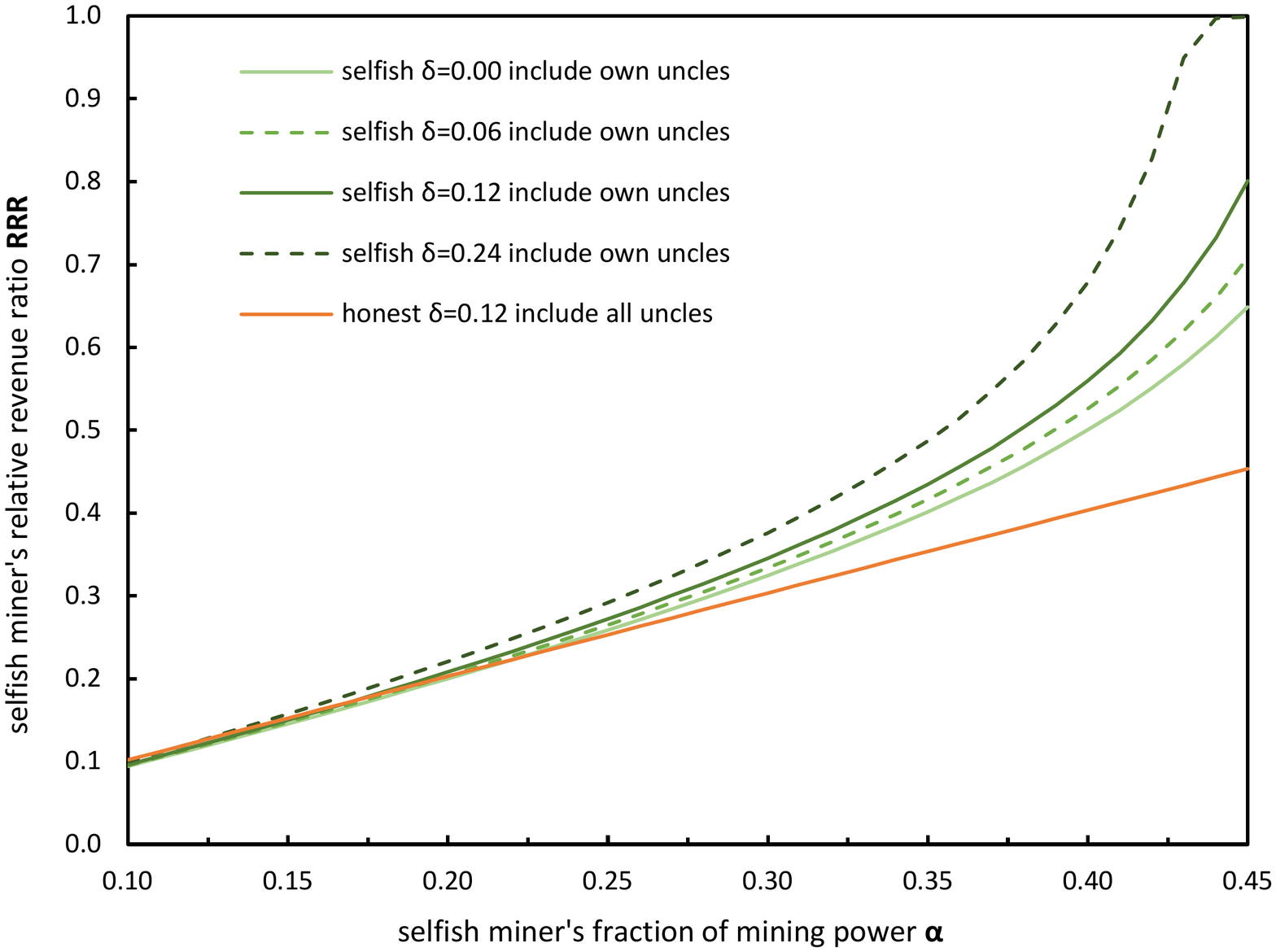}
\label{fig:RRR_1}
\end{subfigure}\hfill
\begin{subfigure}{\columnwidth}
\includegraphics[trim=2cm 1.5cm 2cm 1.5cm, clip=true, width=\columnwidth]{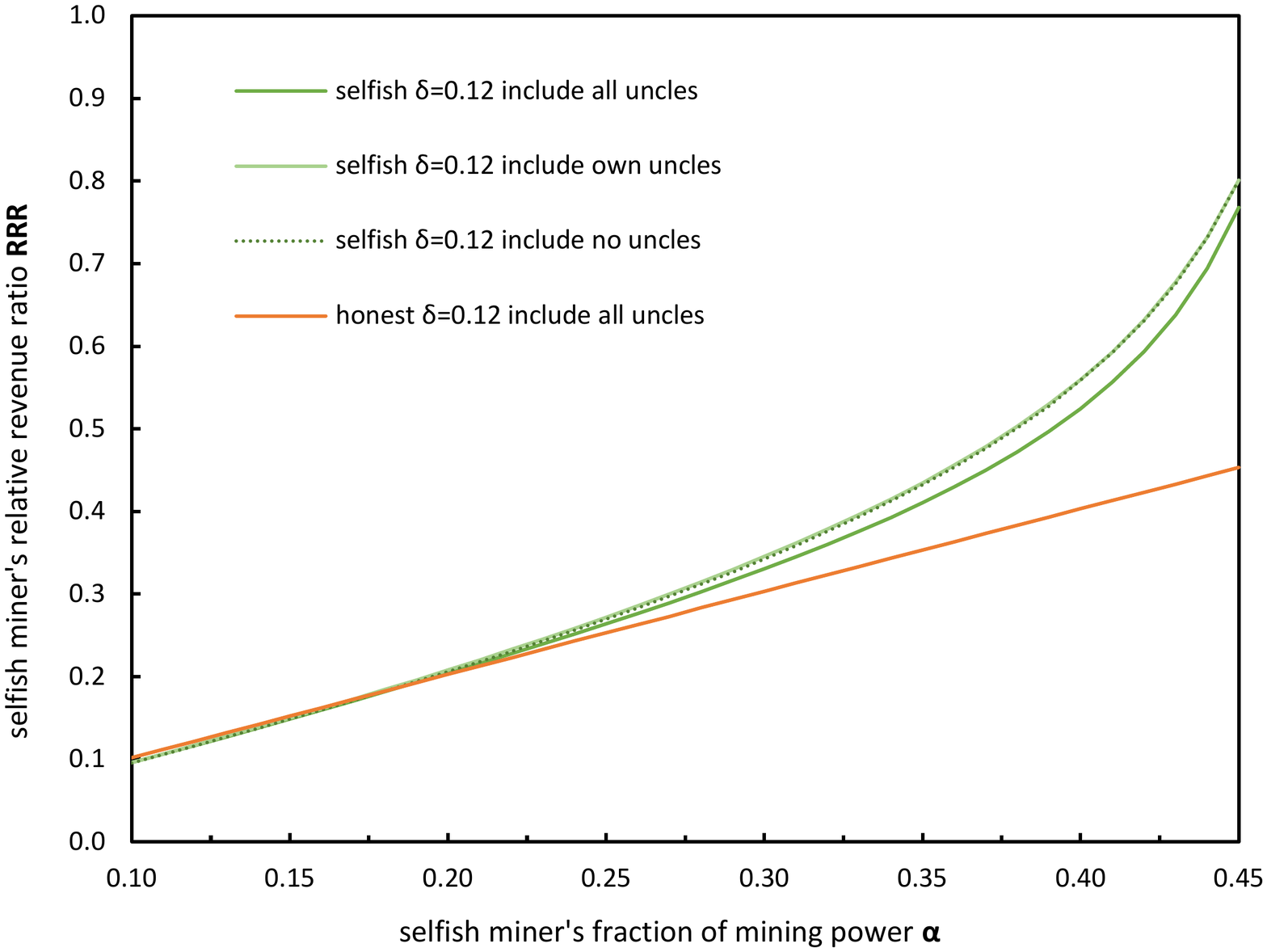}
\label{fig:RRR_2}
\end{subfigure}\hfill
\caption{Simulation results of a selfish miner's $\mathit{RRR}$ (relative revenue ratio) for different values of $\delta$ (honest network's stale block ratio) on the left, simulation results of $\mathit{RRR}$ for a selfish miner that follows different uncle block inclusion strategies on the right.}
\label{fig:RRR}
\end{figure*}

\begin{figure*}
\centering
\begin{subfigure}{\columnwidth}
\includegraphics[trim=2cm 1.5cm 2cm 1.5cm, width=\columnwidth]{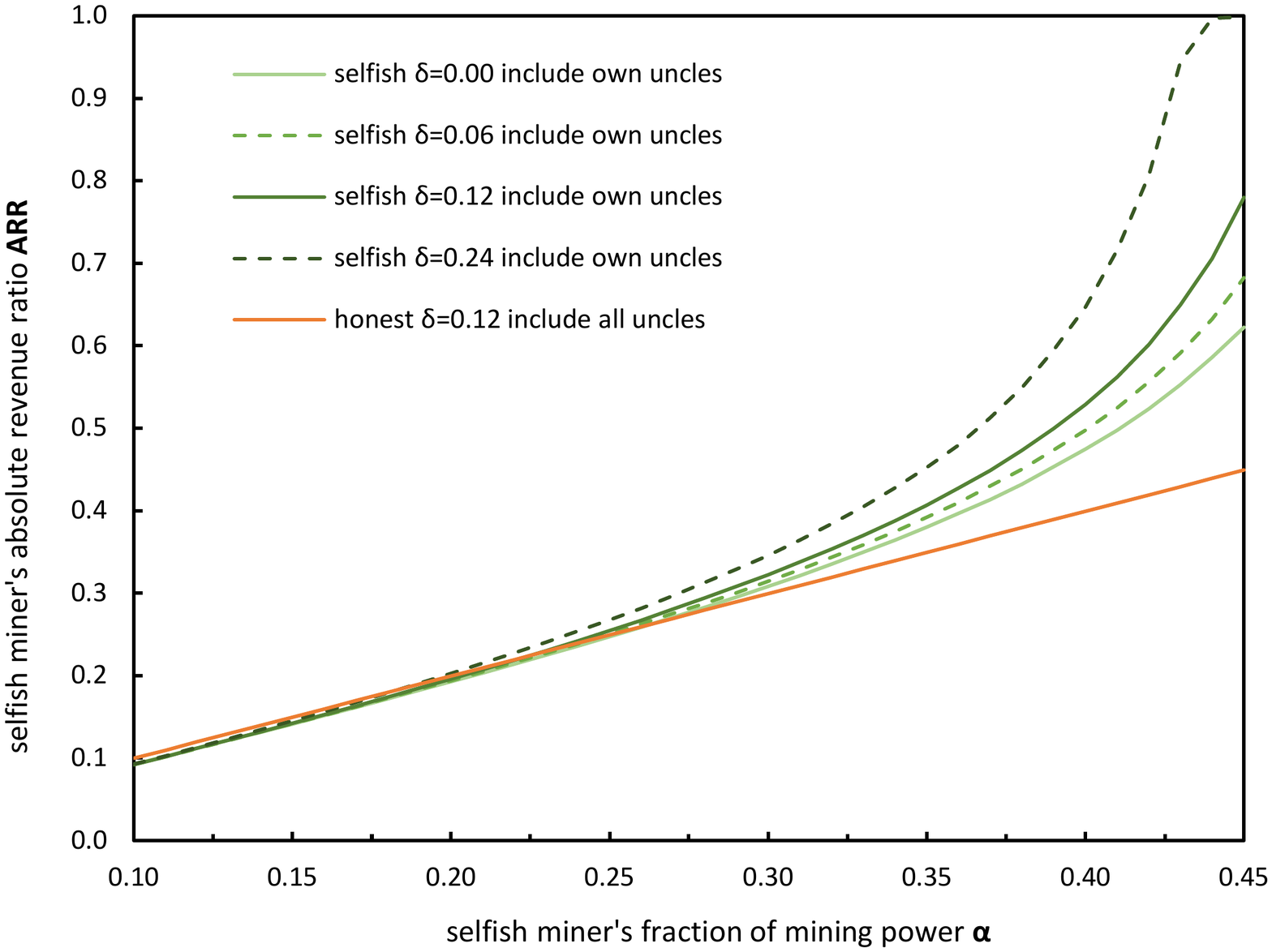}
\label{fig:ARR_1}
\end{subfigure}\hfill
\begin{subfigure}{\columnwidth}
\includegraphics[trim=2cm 1.5cm 2cm 1.5cm, width=\columnwidth]{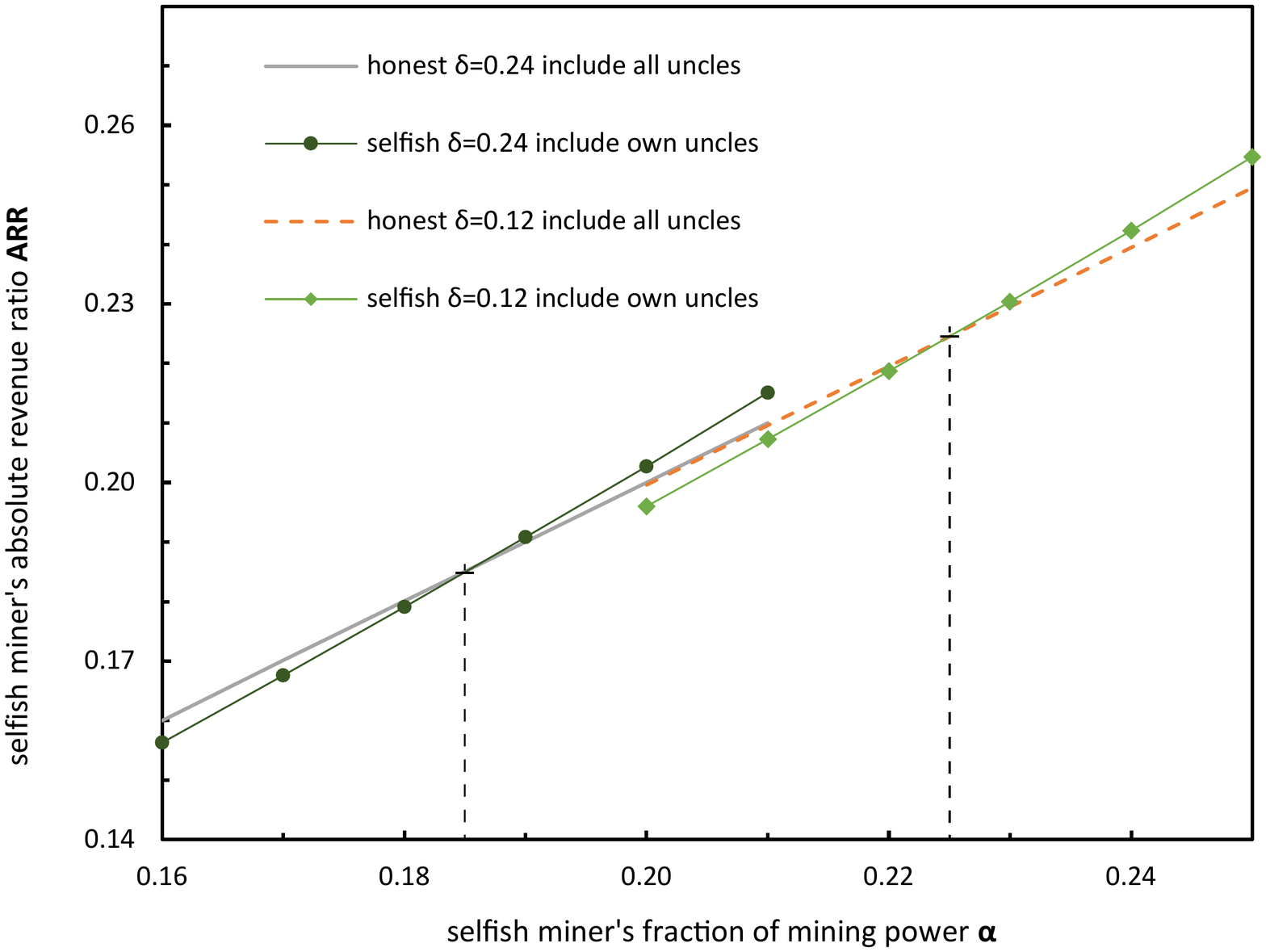}
\label{fig:ARR_2}
\end{subfigure}\hfill
\caption{Simulation results of a selfish miner's $\mathit{ARR}$ (absolute revenue ratio) for different values of $\delta$ (honest network's stale block ratio) on the left, an extract of $\mathit{ARR}$ focusing on the break even of profitability for selfish mining with $\delta = 0.12$ and $\delta = 0.24$ on the right.}
\label{fig:ARR}
\end{figure*}

\begin{figure*}
\centering
\begin{subfigure}{\columnwidth}
\includegraphics[trim=2cm 1.5cm 2cm 1.5cm, width=\columnwidth]{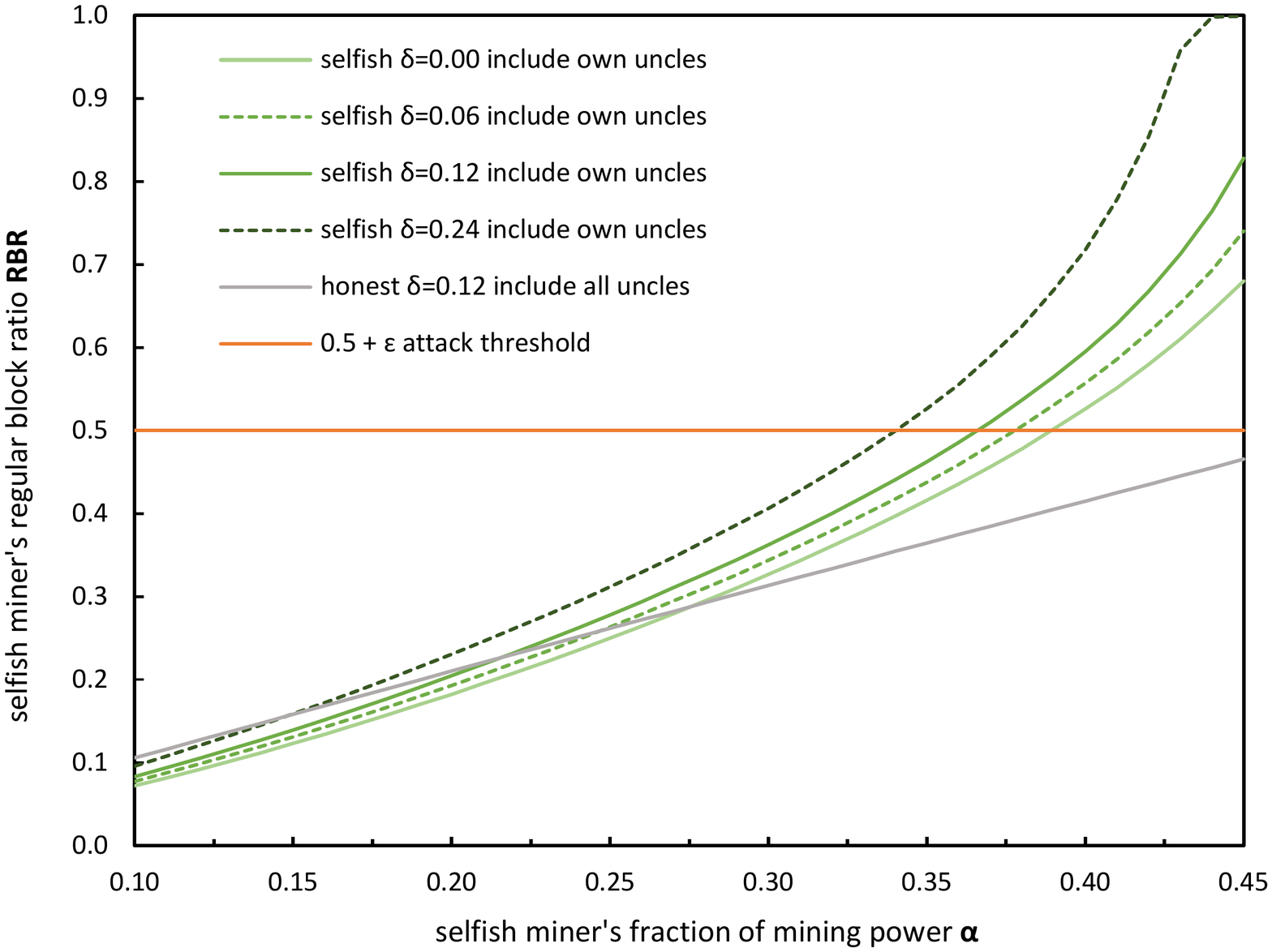}
%\caption{RBR}
\label{fig:RBR}
\end{subfigure}\hfill
\begin{subfigure}{\columnwidth}
\includegraphics[trim=2cm 1.5cm 2cm 1.5cm, width=\columnwidth]{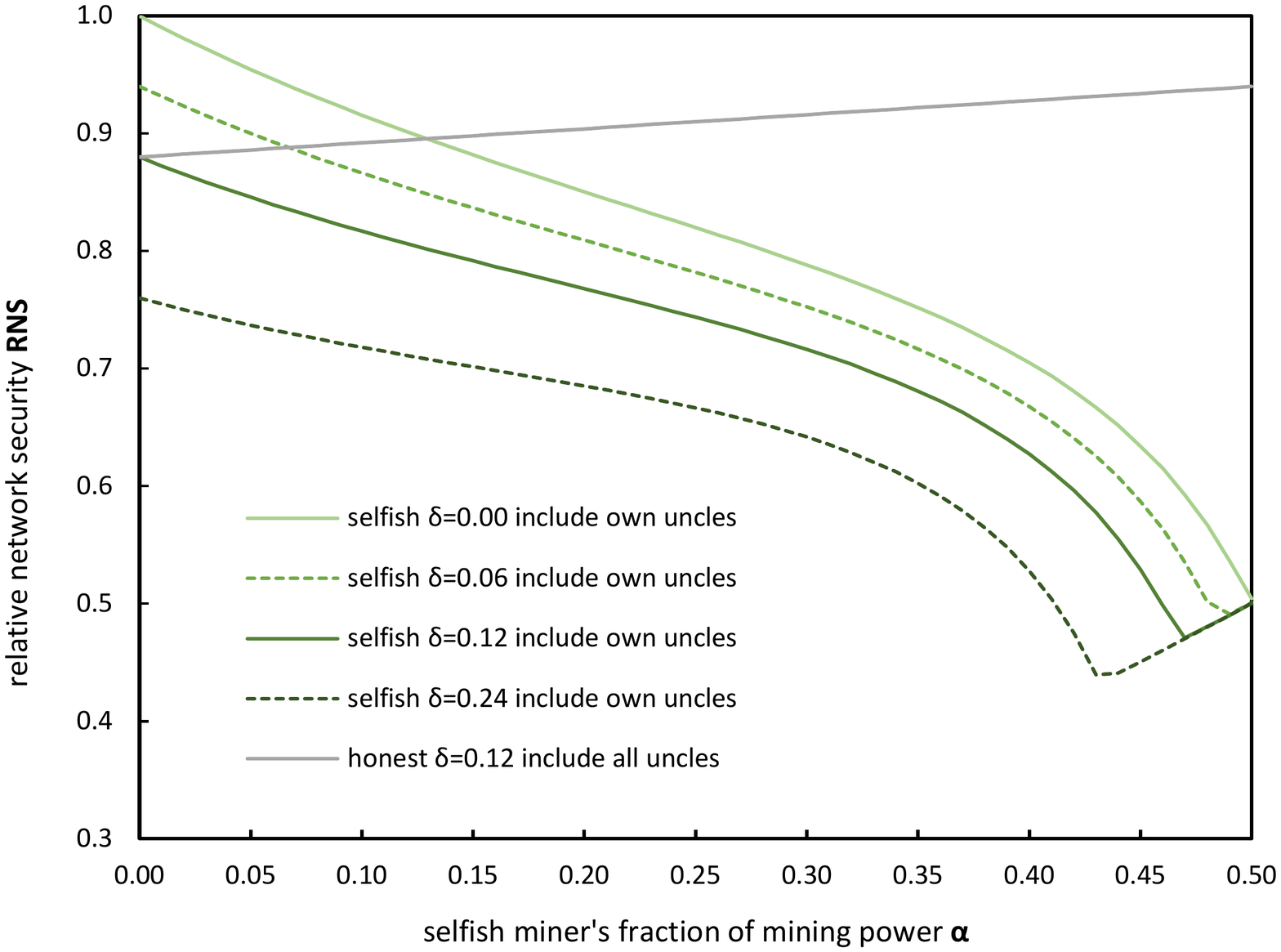}
%\caption{RNS}
\label{fig:RNS}
\end{subfigure}\hfill
\caption{Simulation results of a selfish miner's $\mathit{RBR}$ (regular block ratio) on the left and simulation results of $\mathit{RNS}$ (relative network security) on the right, each for different values of $\delta$ (honest network's stale block ratio).}
\label{fig:RBR-RNS}
\end{figure*}

First of all, a selfish miner benefits from the concept of uncle block inclusion. His relative and absolute revenue increases at higher honest network stale block rates $\delta$ and decreases if he chooses to include the network's stale blocks. This may be explained as follows: A selfish miner creates a stale block every time he looses a race to break a tie against the honest network. With uncle block inclusion, these stale blocks may be referenced afterwards, granting him a partial reward. As $\delta$ increases, the chance of the public chain growing faster than the selfish miner's secret fork decreases. This makes it easier for a selfish miner to outrun the honest network. In case the selfish miner is capable of creating a long fork, the honest network is not capable of referencing the first block of their outrun chain later since it is too far away. But the selfish miner does not profit if he references that block since the average uncle block inclusion reward awarded to him is much lower than the reward for the honest network as the block's miner. Therefore the revenue gap between the selfish miner's uncle block inclusion strategies becomes noteworthy only at higher rates of computational power. An overall positive effect of stale blocks generated by the honest miners to the selfish miner's revenue ratio was also observed by Eyal and Sirer~\cite{bitcoin-selfish-mining} and following work.

Secondly, a selfish miner causes an increased stale block ratio which decreases the average block reward. While the trends of $\mathit{ARR}$ in Fig. \ref{fig:ARR} are similar to those of $\mathit{RRR}$ in Fig. \ref{fig:RRR}, the values of $\mathit{ARR}$ are consistently lower. With blocks yielding only partial reward in Ethereum, $\mathit{RRR}$ is insufficient to estimate the point at which selfish mining becomes profitable. Measured with $\mathit{ARR}$, selfish mining in Ethereum becomes profitable at $\alpha=0.225$ if $\delta=0.12$ and at $\alpha=0.185$ if $\delta=0.24$ (see Fig. \ref{fig:ARR} on the right). For $\delta=0.06$ the break even of profitability is at $\alpha=0.245$.

Lastly, the network's resilience against further attacks is significantly lowered. This is a consequence of the increased stale block ratio: the higher the stale block ratio, the slower the growth of the honest network's main chain. If the selfish miner aims to control the blockchain, he might need only the computational power of $\alpha=0.34$ to overcome the honest network if $\delta=0.24$ as depicted in Fig. \ref{fig:RBR-RNS} on the left. But even if the selfish miner himself does not aim to control all blocks, any other attacker would need less computational power to overcome both the selfish miner and the honest network as quantified by $\mathit{RNS}$ in Fig. \ref{fig:RBR-RNS} on the right. Note that $\mathit{RNS}$ will increase noticeably again after the selfish miner overpowers the honest network because he will create less stale blocks as his chance to loose a race to break a tie diminishes.

%----------------------------------------------------------------------------------------
\section{Consequences}
%----------------------------------------------------------------------------------------
\begin{figure*}
\includegraphics[width=\textwidth]{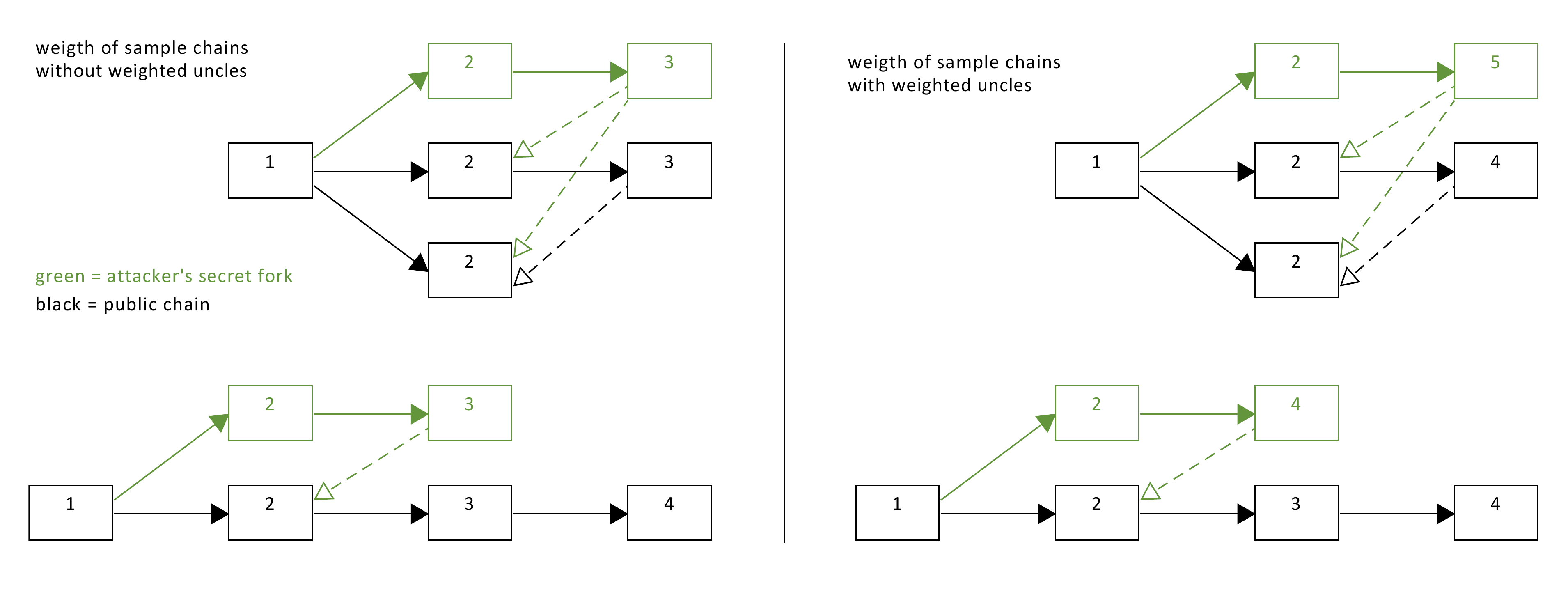}
\caption{Two possible blockchain forks with and without weighted uncle blocks according to ECIP-1029~\cite{ECIP-1029}.}
\label{fig:ecip-1029}
\end{figure*}

As previously shown, a selfish miner benefits from the concept of uncle blocks and can choose to include only own uncle blocks or no uncle blocks at all to further enlarge the honest network’s stale block ratio, both improving his renevue rate. Thus, it may look like a promising idea to weigh uncle blocks during the selection of the best chain similar to the GHOST-rule~\cite{secure-high-tps-bitcoin}. A selfish miner not including uncle blocks would then loose their weight, too. A recent Ethereum Classic Improvement Proposal (ECIP 1029~\cite{ECIP-1029}) suggested to do so.

But because Ethereum uncle blocks need to be actively referenced by some other block, a selfish miner maintaining a secret fork might further profit from being able to also reference blocks previously not broadcast to the honest network. When he releases his secret fork in which he referenced one or more public blocks as uncle blocks, they one-sidedly increase the selfish miner's fork's weight. This means, that even when the selfish miner's fork and the honest network's chain have the same length, the selfish miner's fork is heavier. The following Fig. \ref{fig:ecip-1029} depicts two such situations. The honest network can not reference the first block of the selfish miner's fork until the selfish miner releases it. If moreover the fork is longer than 6 blocks, the honest network can not reference its beginning even after the publication. In such situations, all blocks on the honest network's chain which are not referenced as uncles by the selfish miner become stale blocks. For the heaviest subtree consensus rule to work as intended, Ethereum would have to also change the referencing of uncle blocks to implement the complete GHOST-rule \cite{secure-high-tps-bitcoin}, including its way of following the chains from the root to determine uncle blocks. An analysis of such a fundamental change on the design of Ethereum\cite{12-second-block-time} is beyond the scope of this paper.

%----------------------------------------------------------------------------------------
\section{Related Work}
%----------------------------------------------------------------------------------------
Attacks on the network- or the application-layer~\cite{ethereum-smart-contract-attacks} of Ethereum are not subject of this paper. It does not analyze the so called \emph{uncle mining}~\cite{uncle-mining} as well. Uncle mining exploits the difficulty calculation as before the \emph{Byzantinum} hard fork of Ethereum in October 2017~\cite{EIP-100} by solely creating uncles on purpose. This lowers difficulty on the long term and increases block generation rate which may overcompensate the uncle miner's loss caused by every uncle yielding less mining reward than regular blocks.

Instead, this paper focuses on the original selfish mining strategy as developed by Eyal and Sirer~\cite{bitcoin-selfish-mining}. Curtois and Bahack \cite{subversive-strategies-bitcoin} discussed this attack and criticized it as purely academic and impracticable. Nevertheless, Sapirshtein et al.~\cite{optimal-selfish-mining-bitcoin} explored the $\epsilon$-optimal policies for when a selfish miner in Bitcoin should release his secret fork, based on an extended model of Eyal and Sirer~\cite{bitcoin-selfish-mining} and an optimization algorithm. In parallel, Nayak et al.~\cite{stubborn-mining} enhanced the attack by introducing stubborn mining, in which the selfish miner also has more options when to release his secret fork and combined these with additional network layer attacks. Altogether, Sapirshtein et al.~\cite{optimal-selfish-mining-bitcoin} and Nayak et al.~\cite{stubborn-mining} significantly improved the profitability of selfish mining.

This paper does not aim to optimize selfish mining in Ethereum. Instead, it quantifies how the concept of uncles generally influences the profitability of selfish mining. Therefore only the basic selfish mining strategy was adopted but combined with further Ethereum-specific variables such as the network stale block ratio, causing network lag.

Previous work by Decker and Wattenhofer~\cite{information-propagation-bitcoin} pointed out how block size affects propagation times and that network lag is one reason for blockchain forks. An increased vulnerability to $0.5+\epsilon$ attacks was concluded, but not quantified. The following work of Sompolinsky and Zohar~\cite{secure-high-tps-bitcoin} quantified a security threshold against the $0.5+\epsilon$ attack in relation to the transaction throughput and developed the advanced consensus rule GHOST to overcome this issue. The present paper tries to quantify degradation of security against all possible attacks.

Gervais et al.~\cite{security-performance-blockchain} quantified the general trade-offs between network throughput and blockchain security. Kiayias and Panagiotakos~\cite{trees-chains-tps-blockchain} proved that blockchain security may be preserved at high transaction throughput by using GHOST. This paper shows that the adaptation of GHOST as proposed for Ethereum classic in \cite{ECIP-1029} would actually weaken the blockchain security and make Ethereum classic more vulnerable to selfish attacks.

Luu et al.~\cite{power-splitting-games} perform a game theoretic analysis of the security of Bitcoin mining pool protocols against block withholding. It focuses on mining profitability but does not quantify the effects on the blockchain's security. The attack is orthogonal to the selfish mining attack of the present paper.

%----------------------------------------------------------------------------------------
\section{Conclusion}
%----------------------------------------------------------------------------------------
Compared to selfish mining in Bitcoin \cite{bitcoin-selfish-mining}, the concept of rewards for uncle blocks, as done in Ethereum, lowers the amount of computational power at which selfish mining becomes a viable strategy. While in Bitcoin the threshold for $\gamma=0.5$ is 0.25, the threshold in Ethereum may be as low as 0.185 depending on the honest network's stale block ratio. Optimized strategies possibly lower this threshold even further. Moreover, the overall network's resilience against other attacks -- such as double spending -- is significantly lowered during selfish mining caused by a high stale block ratio. Any attacker would then need less mining power to outrun the overall network temporarily or even entirely. The adaption of optimal selfish mining strategies as well as the in-depth-analysis of countermeasures is left for future work.

%----------------------------------------------------------------------------------------
\section*{Acknowledgments}
%----------------------------------------------------------------------------------------
We are grateful to Jeremy Clark, Felix Merkl, Bartosz Golis and Diana Scheuring as well as the anonymous reviewers for their valuable advice and discussions.

%----------------------------------------------------------------------------------------
\bibliography{refs}
%----------------------------------------------------------------------------------------

\end{document}